\documentclass[12pt,dviwindows]{article}

\usepackage{color,epsf,graphicx}
\usepackage{amsmath,amssymb}
\usepackage{psfrag}
\usepackage[colorlinks=true,citecolor=blue,linkcolor=red]{hyperref}

\newcounter{comment}

{\refstepcounter{comment}%
\begin{quote}
\ttfamily\small$\blacksquare$ \textbf{\underline{Comment} $\sharp$\thecomment:}}%
{\end{quote}}

{
\begin{quote}
\ttfamily\small$\blacktriangleright$ \textbf{\underline{Reply} $\sharp$\thecomment:}}%
{\end{quote}}

\font\cmss=cmss12 
\def\1{\hbox{{1}\kern-.25em\hbox{l}}}
\def\bfZ{\relax{\hbox{\cmss Z\kern-.4em Z}}}

\def\ru1{\rule[-0.4truecm]{0mm}{1truecm}}

\renewcommand{\thefootnote}{\fnsymbol{footnote}}
\renewcommand{\bar}[1]{\overline{#1}}

\def\ru1{\rule[-0.4truecm]{0mm}{1truecm}}
\def\upleftarrow#1{\overleftarrow{#1}}
\def\uprightarrow#1{\overrightarrow{#1}}
\def\thru#1{\mathrel{\mathop{#1\!\!\!/}}}
\def\be{\begin{eqnarray}}
\def\ee{\end{eqnarray}}
\def\bea{\begin{eqnarray}}
\def\eea{\end{eqnarray}}

\def\D2{{\bf \Delta}_\perp^2}
\def\0T{{\bf 0}_\perp}

\begin{document}



\setcounter{footnote}{0}
\renewcommand{\thefootnote}{\fnsymbol{footnote}}
\renewcommand{\bar}[1]{\overline{#1}}
\newcommand{\ie}{{\it i.e.}}
\newcommand{\eg}{{\it e.g.,}}
\newcommand{\btt}[1]{{\tt$\backslash$#1}}
\newcommand{\half}{{$\frac{1}{2}$}} 
\newcommand{\ket}[1]{\left\vert\,{#1}\right\rangle}
\newcommand{\VEV}[1]{\left\langle{#1}\right\rangle}

\def\ru1{\rule[-0.4truecm]{0mm}{1truecm}}
\def\upleftarrow#1{\overleftarrow{#1}}
\def\uprightarrow#1{\overrightarrow{#1}}
\def\thru#1{\mathrel{\mathop{#1\!\!\!/}}}

\def\senk#1{\bbox{#1}_\perp}
\def\ha{{1\over 2}}
\def\ub#1{\underline{#1}}
\def\ths{\thinspace}
\def\psibar{\overline{\psi}}
\def\del{\partial}
\def\ra{\rightarrow}
\def\eg{{\it e.g.}}
\def\g{\gamma}

\newpage
\begin{flushright}
March 2010
\end{flushright}

\bigskip

\begin{center}
{\Large
\bf Light-Cone Wavefunction Representations of\\
Sivers and Boer-Mulders Distribution Functions}\\


\vspace{1.0cm}

\centerline{{\bf Dae Sung Hwang}}


\vspace{4mm} \centerline{\it Department of Physics, Sejong
University, Seoul 143--747, South Korea}

\vspace{2.5cm}

{\bf Abstract}\\
\end{center}
\noindent
We find the light-cone wavefunction representations of
the Sivers and Boer-Mulders distribution functions.
A necessary condition for the existence of these representations is that
the light-cone wavefunctions have complex phases.
We induce the complex phases by incorporating the final-state interactions
into the light-cone wavefunctions.
For the scalar and axial-vector diquark models for nucleon,
we calculate explicitly the Sivers and Boer-Mulders distribution
functions from the light-cone wavefunction representations.
We obtain the results that the Sivers distribution function has the opposite signs
with the factor 3 difference in magnitude for the two models,
whereas the Boer-Mulders distribution function has the same
sign and magnitude.
We can understand these results from the properties of the
light-cone wavefunction representations of the Sivers and Boer-Mulders
distribution functions.
\\

\vfill


\thispagestyle{empty}
\pagebreak

\setlength{\baselineskip}{13pt}

\section{Introduction}

\begin{figure}[h!]
\begin{center}
\includegraphics[width=0.75\columnwidth]{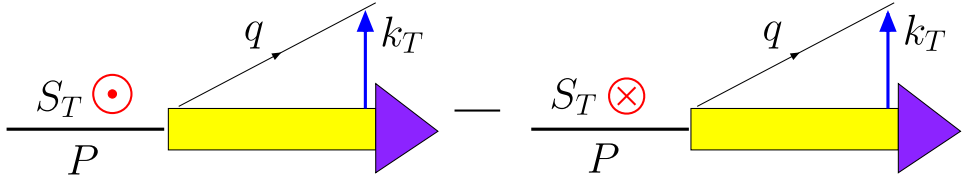}\\
\caption{\it Schematic depiction of the Sivers distribution function
$f_{1T}^{\perp}$. The spin vector $S_T$ of the nucleon points out of and
into the page, respectively, and $k_T$ is the transverse momentum of the extracted quark.}
\label{fig:siversdiagram}
\end{center}
\end{figure}

\begin{figure}[h!]
\begin{center}
\includegraphics[width=0.75\columnwidth]{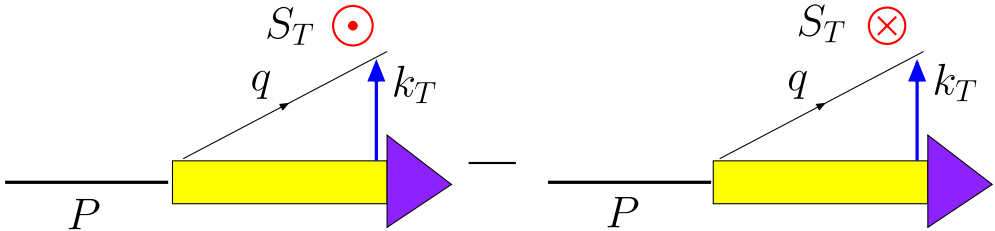}\\
\caption{\it Schematic depiction of the Boer-Mulders distribution function
$h_{1}^{\perp}$. The spin vector $S_T$ of the quark points out of and
into the page, respectively, and $k_T$ is the transverse momentum of the extracted quark.}
\label{fig:BMdiagram}
\end{center}
\end{figure}

It was found that the final-state interaction of quark and gluon
induces the single-spin
asymmetry in the semi-inclusive deep inelastic scattering at the
twist-two level \cite{BHS}.
Then, this time-odd twist-two effect was interpreted as the Sivers
effect \cite{Sivers} by finding that the final-state interaction can be treated as
the source of the time-odd Sivers distribution function
\cite{Collins02,JY,BJY,BBH,BMP}.
It is also often referred to as ``naively $T$-odd'',
because the appearance of this function does not imply a violation of
time-reversal invariance, since they can arise through the final-state
interactions.
With these developments, the existence of the Sivers distribution function
has gained a firm theoretical support.
The Sivers distribution function $f_{1T}^{\perp}$
describes the difference between the momentum distributions of unpolarized quark
inside the nucleons transversely polarized in opposite directions.
There is another quark distribution function of the nucleon induced by
the final-state interaction of quark and gluon, which is called
the Boer-Mulders distribution function $h_{1}^{\perp}$ \cite{BM98}.
$h_{1}^{\perp}$ describes the difference between the momentum distributions
of the quarks transversely polarized in opposite directions inside
unpolarized nucleon.
The distribution functions $f_{1T}^{\perp}$ and $h_{1}^{\perp}$ are depicted
in Figs. \ref{fig:siversdiagram} and \ref{fig:BMdiagram}.

\vspace{-0.8cm}

The light-cone wavefunctions are valuable for studying the hadronic
processes by treating the non-perturbative effects in a relativistically
covariant way \cite{BL,BPP}.
The formulas which express the electromagnetic form factors and
the generalized parton distribution functions in terms of the light-cone
wavefunctions were found
in Refs. \cite{BD80,BH99,BHMS} and Refs. \cite{DFJK,BDH}, respectively.
In Ref. \cite{BGH06} the light-cone wavefunction representation of
the nucleon electric dipole moment was found by introducing the complex phases of
the light-cone wavefunctions, and studied a general relation connecting
nucleon electric dipole and anomalous magnetic moments.

In this paper we derive the formulas which express the Sivers and
Boer-Mulders distribution functions in terms of the matrix elements of
the nucleon spin states.
Then we derive the light-cone wavefunction representations of
the Sivers and Boer-Mulders distribution functions,
and we calculate these functions for the scalar and axial-vector diquark models by using these
representations.

The light-cone wavefunction representations show that
the Sivers distribution function is given by the overlap of the wavefunctions of the same
quark spin states of the opposite nucleon spin states,
whereas the Boer-Mulders distribution function
is given by the overlap of the wavefunctions of the opposite quark spin states
within a given nucleon spin state.
From these properties of their light-cone wavefunction representations,
we can understand why the Sivers distribution function has the opposite signs
with the factor 3 difference in magnitude for the scalar and axial-vector diquark models,
whereas the Boer-Mulders distribution function has the same sign
and magnitude for these diquark models.

\section{Sivers and Boer-Mulders Distribution Functions}

The $k_T$-dependent unpolarized quark distribution function $f_1(x,{\vec k}_{\perp})$,
the Sivers distribution function $f_{1T}^{\perp}(x,{\vec k}_{\perp})$
and the Boer-Mulders distribution function $h_1^{\perp}(x,{\vec k}_{\perp})$ are parts
of the proton correlation function $\Phi(x,{\vec k}_{\perp}:P,S)$ \cite{BM98}:
\begin{equation}
\Phi(x,{\vec k}_{\perp}:P,S)={M\over 2P^+}\Big[
f_1(x,{\vec k}_{\perp}){\gamma\cdot P\over M}
+f_{1T}^{\perp}(x,{\vec k}_{\perp})\epsilon_{\mu\nu\rho\sigma}
{\gamma^{\mu}P^\nu k_{\perp}^\rho S_T^\sigma \over M^2}
+h_1^{\perp}(x,{\vec k}_{\perp}){{\sigma}_{\mu\nu}k_{\perp}^{\mu}P^{\nu} \over M^2}
+\cdots \Big]\ ,
\label{crf}
\end{equation}
from which we find that
$f_1(x,{\vec k}_{\perp})$ and $f_{1T}^{\perp}(x,{\vec k}_{\perp})$
can be defined through matrix elements
of the bilinear vector current:
\begin{eqnarray}
&&\int\frac{d y^-d^2{\vec y}_{\perp}}{16\pi^3}\;
e^{ix P^+y^--i{\vec k}_{\perp}\cdot {\vec y}_{\perp}}\;
\langle P,{\vec S}_{\perp} | \bar\psi(0)\,\gamma^+\,\psi(y)\,|P,{\vec S}_{\perp}
\rangle
\Big|_{y^+=0}
\label{siv1}\\
&=&
{1\over 2 P^+}\ \Big[\ f_1(x,{\vec k}_{\perp})\ {\bar U}(P,{\vec S}_{\perp})
\ {\gamma^+} \ U(P,{\vec S}_{\perp})
\ +\
f_{1T}^{\perp}(x,{\vec k}_{\perp})\ {k_{\perp}^i\over M}\ {\bar U}(P,{\vec S}_{\perp})
\ \sigma^{i+} \ U(P,{\vec S}_{\perp}) \ \Big]
\ ,
\nonumber
\end{eqnarray}
where
\begin{equation}
{1\over 2 P^+}
{\bar U}(P,{\vec S}_{\perp}) \sigma^{i+} U(P,{\vec S}_{\perp})= 
\epsilon^{ji}S_{\perp}^j\ \qquad
{\rm with}\ \epsilon^{12}=-\epsilon^{21}=1\ .
\label{siv2}
\end{equation}

For an explicit calculation, let us consider the case of
${\vec S}_{\perp}=(S_{\perp}^1,S_{\perp}^2)=(0,1)$ for the
transverse spin in (\ref{siv1}).
Then, the proton state is given by
$(|P,\uparrow \rangle +i |P,\downarrow \rangle)/{\sqrt{2}}$
and Eq. (\ref{siv1}) becomes
\begin{equation}
{\cal B}\ {\langle P,\uparrow | -i \langle P,\downarrow | \over {\sqrt{2}}}
\bar\psi(0)\,\gamma^+\,\psi(y)\,
{|P,\uparrow \rangle +i |P,\downarrow \rangle \over {\sqrt{2}}}
\Big|_{y^+=0}
=f_1(x,{\vec k}_{\perp})\
\ -\ S_{\perp}^2\ {k_{\perp}^1\over M}\ f_{1T}^{\perp}(x,{\vec k}_{\perp})
\ ,
\label{siv3ab}
\end{equation}
where
\begin{equation}
{\cal B}\ \equiv \ \int\frac{d y^-d^2{\vec y}_{\perp}}{16\pi^3}\;
e^{ix P^+y^--i{\vec k}_{\perp}\cdot {\vec y}_{\perp}}\ .
\label{caladef}
\end{equation}
From (\ref{siv3ab}) we have
\begin{eqnarray}
f_1(x,{\vec k}_{\perp})&=&
{\cal B}\ 
{1\over 2}\;\Big[\langle P,\uparrow |J^+(y)|P,\uparrow \rangle 
+ \langle P,\downarrow |J^+(y)|P,\downarrow \rangle \Big]
\Big|_{y^+=0} \ ,
\label{siv4}\\
-{k_{\perp}^1\over M}\ f_{1T}^{\perp}(x,{\vec k}_{\perp})&=&
{\cal B}\ 
{i\over 2}\;\Big[\langle P,\uparrow |J^+(y)|P,\downarrow \rangle 
- \langle P,\downarrow |J^+(y)|P,\uparrow \rangle \Big]
\Big|_{y^+=0}
\ ,
\qquad
\label{siv5}
\end{eqnarray}
where
$J^+(y)=\bar\psi(0)\,\gamma^+\,\psi(y)$.

On the other hand, from (\ref{crf}) the Boer-Mulders distribution function
$h_{1}^{\perp}(x,{\vec k}_{\perp})$ can be defined through matrix elements
of the bilinear tensor current:
\begin{eqnarray}
&&\int\frac{d y^-d^2{\vec y}_{\perp}}{16\pi^3}\;
e^{ix P^+y^--i{\vec k}_{\perp}\cdot {\vec y}_{\perp}}\;
\langle P,{\vec S}_{\perp} | \, \bar\psi(0)\,\sigma^{i+}\,\psi(y)\,
|P,{\vec S}_{\perp}
\rangle
\Big|_{y^+=0}
\label{siv1h1p}\\
&&\qquad =\
{1\over 2 P^+}\ \Big[\ h_{1}^{\perp}(x,{\vec k}_{\perp})\ 
{k_{\perp}^i\over M}\ {\bar U}(P,{\vec S}_{\perp})
\ \gamma^{+} \ U(P,{\vec S}_{\perp}) \ \Big]
\ ,
\nonumber
\end{eqnarray}
which gives
\begin{equation}
{k_{\perp}^i\over M}\ h_{1}^{\perp}(x,{\vec k}_{\perp})
\ =\ {1\over 2}\
{\cal B}\
\Big( \Big[\langle P,\uparrow |\, \bar\psi(0)\,\sigma^{i+}\,\psi(y)\,
|P,\uparrow \rangle \Big]
+
\Big[\langle P,\downarrow |\, \bar\psi(0)\,\sigma^{i+}\,\psi(y)\,
|P,\downarrow \rangle \Big] \Big)
\Big|_{y^+=0}
\ .
\label{hp1a}
\end{equation}

\section{Light-Cone Wavefunction Representations of Sivers and Boer-Mulders
Functions}


The expansion of the proton
eigensolution $\ket{\psi_p}$ on the eigenstates $\{\ket{n} \}$ of the
free Hamiltonian $H_{LC}$ gives the light-cone Fock expansion:
\begin{eqnarray}
\left\vert \psi_p(P^+, {\vec P_\perp} )\right> &=& \sum_{n}\
\prod_{i=1}^{n}
{{\rm d}x_i\, {\rm d}^2 {\vec k_{\perp i}}
\over \sqrt{x_i}\, 16\pi^3}\ \,
16\pi^3 \delta\left(1-\sum_{i=1}^{n} x_i\right)\,
\delta^{(2)}\left(\sum_{i=1}^{n} {\vec k_{\perp i}}\right)
\label{a318}
\\
&& \qquad \rule{0pt}{4.5ex}
\times \psi_n(x_i,{\vec k_{\perp i}},
\lambda_i) \left\vert n;\,
x_i P^+, x_i {\vec P_\perp} + {\vec k_{\perp i}}, \lambda_i\right>.
\nonumber
\end{eqnarray}
The plus component momentum fractions $x_i = k^+_i/P^+$ and the transverse
momenta ${\vec k_{\perp i}}$ of partons represent the relative momentum
coordinates of the light-cone wavefunctions.
The physical transverse momenta of partons are ${\vec p_{\perp i}}
= x_i {\vec P_\perp} + {\vec k_{\perp i}}.$ The $\lambda_i$ label the
light-cone spin projections of the partons along the
quantization direction $z$. The $n$-particle states are normalized as
\begin{equation}
\left< n;\, p'_i{}^+, {\vec p\,'_{\perp i}}, \lambda'_i \right. \,
\left\vert n;\,
p^{~}_i{}^{\!\!+}, {\vec p^{~}_{\perp i}}, \lambda_i\right>
= \prod_{i=1}^n 16\pi^3
p_i^+ \delta(p'_i{}^{+} - p^{~}_i{}^{\!\!+})\
\delta^{(2)}( {\vec p\,'_{\perp i}} - {\vec p^{~}_{\perp i}})\
\delta_{\lambda'_i\, \lambda^{~}_i}\ .
\label{normalize}
\end{equation}

{}From (\ref{siv4}) and (\ref{siv5}) we get
\begin{equation}
f_1(x,{\vec k}_{\perp})
=
{\cal C}\
{1\over 2}\ \Big[ \psi^{\uparrow \ *}_{(n)}(x_i,
  {\vec{k}}_{\perp i},\lambda_i) \
\psi^{\uparrow}_{(n)}(x_i, {\vec{k}}_{\perp i},\lambda_i)
\ +\
\psi^{\downarrow \ *}_{(n)}(x_i,
  {\vec{k}}_{\perp i},\lambda_i) \
\psi^{\downarrow}_{(n)}(x_i, {\vec{k}}_{\perp i},\lambda_i)
\Big] \ ,
\label{p11sa}
\end{equation}
\begin{equation}
-{k_{\perp}^1\over M}\ f_{1T}^{\perp}(x,{\vec k}_{\perp})
=
{\cal C}\
{i\over 2}\ \Big[ \psi^{\uparrow \ *}_{(n)}(x_i,
  {\vec{k}}_{\perp i},\lambda_i) \
\psi^{\downarrow}_{(n)}(x_i, {\vec{k}}_{\perp i},\lambda_i)
\ -\
\psi^{\downarrow \ *}_{(n)}(x_i,
  {\vec{k}}_{\perp i},\lambda_i) \
\psi^{\uparrow}_{(n)}(x_i, {\vec{k}}_{\perp i},\lambda_i)
\Big] \ ,
\label{p11sb}
\end{equation}
where
\begin{equation}
{\cal C}\ \equiv \
\sum_{n, \lambda_i}
\int \prod_{i=1}^{n}
{{\rm d}x_{i}\, {\rm d}^2{\vec{k}}_{\perp i} \over 16\pi^3 }\ \,
16\pi^3 \delta\left(1-\sum_{j=1}^n x_j\right) \, \delta^{(2)}
\left(\sum_{j=1}^n {\vec{k}}_{\perp j}\right)\
\delta(x-x_{1})\ \delta^{(2)}({\vec{k}}_{\perp}-{\vec{k}}_{\perp 1})\ .
\label{pp1}
\end{equation}
As we see in (\ref{p11sb}), the Sivers distribution
function is given by the product of the light-cone wavefunctions which
have opposite proton spin states and same quark spin states.


From (\ref{hp1a}) we have
\begin{eqnarray}
{k_{\perp}^1\over M}\ h_{1}^{\perp}(x,{\vec k}_{\perp})
&=&
{{\cal C}\over 2}\ (-i)\
\Big(
\Big[ \ \psi^{\uparrow \ *}_{(n)}(x_i,
  {\vec{k}}_{\perp i},{\lambda}^{\prime}_1=\downarrow ,\lambda_{i\ne 1})
\ \psi^{\uparrow}_{(n)}(x_i, {\vec{k}}_{\perp i},\lambda_1=\uparrow ,\lambda_{i\ne 1})
\nonumber\\
&&\qquad\ \ \ -\
\psi^{\uparrow \ *}_{(n)}(x_i, 
  {\vec{k}}_{\perp i},{\lambda}^{\prime}_1=\uparrow ,\lambda_{i\ne 1})
\ \psi^{\uparrow}_{(n)}(x_i, {\vec{k}}_{\perp i},\lambda_1=\downarrow ,\lambda_{i\ne 1})
\ \Big]
\nonumber\\
&&\ \
\ \ \ \ \ \ +\
\Big[ \ \psi^{\downarrow \ *}_{(n)}(x_i,
  {\vec{k}}_{\perp i},{\lambda}^{\prime}_1=\downarrow ,\lambda_{i\ne 1})
\ \psi^{\downarrow}_{(n)}(x_i, {\vec{k}}_{\perp i},\lambda_1=\uparrow ,\lambda_{i\ne 1})
\nonumber\\
&&\qquad\ \ \ -\
\psi^{\downarrow \ *}_{(n)}(x_i,
  {\vec{k}}_{\perp i},{\lambda}^{\prime}_1=\uparrow ,\lambda_{i\ne 1})
\ \psi^{\downarrow}_{(n)}(x_i, {\vec{k}}_{\perp i},\lambda_1=\downarrow ,\lambda_{i\ne 1})
\ \Big]
\Big) \ . \ \ \ \ \ \
\label{p11sbh1pan}
\end{eqnarray}
As we see in (\ref{p11sbh1pan}), the Boer-Mulders
distribution function is given by the product of the light-cone wavefunctions which
have same proton spin states and opposite quark spin states,
whereas we found in (\ref{p11sb}) that the Sivers distribution
function is given by the product of the light-cone wavefunctions which
have opposite proton spin states and same quark spin states.


\section{Explicit Calculations in Diquark Models}

\subsection{Scalar Diquark Model}

In this subsection we calculate the Sivers and Boer-Mulders distribution
functions of the scalar diquark model by using the light-cone wavefunction
representations derived in section 3.
In the scalar diquark model,
the $J^z = + {1\over 2}$ two particle Fock state is given by
\cite{BHS,BHMS}
\begin{eqnarray}
&&\left|\Psi^{\uparrow}_{\rm two \ particle}(P^+=1, \vec P_\perp = \vec
0_\perp)\right>
\label{sn1}\\
&=&
\int\frac{{\mathrm d} x \, {\mathrm d}^2
           {\vec k}_{\perp} }{\sqrt{x(1-x)}\, 16 \pi^3}
\Big[ \
\psi^{\uparrow}_{+\frac{1}{2}} (x,{\vec k}_{\perp})\,
\left| +\frac{1}{2}\, ;\,\, x\, ,\,\, {\vec k}_{\perp} \right>
+\psi^{\uparrow}_{-\frac{1}{2}} (x,{\vec k}_{\perp})\,
\left| -\frac{1}{2}\, ;\,\, x\, ,\,\, {\vec k}_{\perp} \right>\ \Big]\ ,
\nonumber
\end{eqnarray}
where
\begin{equation}
\left
\{ \begin{array}{l}
\psi^{\uparrow}_{+\frac{1}{2}} (x,{\vec k}_{\perp})=\frac{(m+xM)}{x}\,
\varphi \ ,\\
\psi^{\uparrow}_{-\frac{1}{2}} (x,{\vec k}_{\perp})=
-\frac{(+k^1+{i} k^2)}{x }\,
\varphi \ .
\end{array}
\right.
\label{sn2}
\end{equation}
The scalar part of the wavefunction $\varphi$ is given by
\begin{equation}
\varphi (x,{\vec k}_{\perp}) = \frac{g}{\sqrt{1-x}}\
\frac{1}{M^2-{{\vec k}_{\perp}^2+m^2 \over x}
-{{\vec k}_{\perp}^2+\lambda^2 \over 1-x}}
=-g{x{\sqrt{1-x}}\over {\vec k}_{\perp}^2+B}
\ ,
\label{wfdenom}
\end{equation}
where
\begin{equation}
B=x(1-x)\Bigl( -M^2+{m^2\over x}+{{\lambda}^2\over 1-x} \Bigr)\ .
\label{wfdenom2}
\end{equation}

Similarly, the $J^z = - {1\over 2}$ two particle Fock state is given by
\begin{eqnarray}
&&\left|\Psi^{\downarrow}_{\rm two \ particle}(P^+=1, \vec P_\perp =
\vec 0_\perp)\right>
\label{sn1a}\\
&=&
\int\frac{{\mathrm d} x \, {\mathrm d}^2
           {\vec k}_{\perp} }{\sqrt{x(1-x)}\, 16 \pi^3}
\Big[ \
\psi^{\downarrow}_{+\frac{1}{2}} (x,{\vec k}_{\perp})\,
\left| +\frac{1}{2}\, ;\,\, x\, ,\,\, {\vec k}_{\perp} \right>
+\psi^{\downarrow}_{-\frac{1}{2}} (x,{\vec k}_{\perp})\,
\left| -\frac{1}{2}\, ;\,\, x\, ,\,\, {\vec k}_{\perp} \right>\ \Big]\ ,
\nonumber
\end{eqnarray}
where
\begin{equation}
\left
\{ \begin{array}{l}
\psi^{\downarrow}_{+\frac{1}{2}} (x,{\vec k}_{\perp})=
-\frac{(-k^1+{i} k^2)}{x }\,
\varphi \ ,\\
\psi^{\downarrow}_{-\frac{1}{2}} (x,{\vec k}_{\perp})=\frac{(m+xM)}{x}\,
\varphi \ .
\end{array}
\right.
\label{sn2a}
\end{equation}
The coefficients of $\varphi$ in Eqs.  (\ref{sn2})
and (\ref{sn2a}) are the matrix elements of
$\frac{\overline{u}(k^+,k^-,{\vec k}_{\perp})}{{\sqrt{k^+}}}
\frac{u (P^+,P^-,{\vec P}_{\perp})}{{\sqrt{P^+}}}$
which are
the numerators of the wavefunctions corresponding to
each constituent spin $s^z$ configuration.

\begin{figure}[h!]
\begin{center}
\includegraphics[width=1.0\columnwidth]{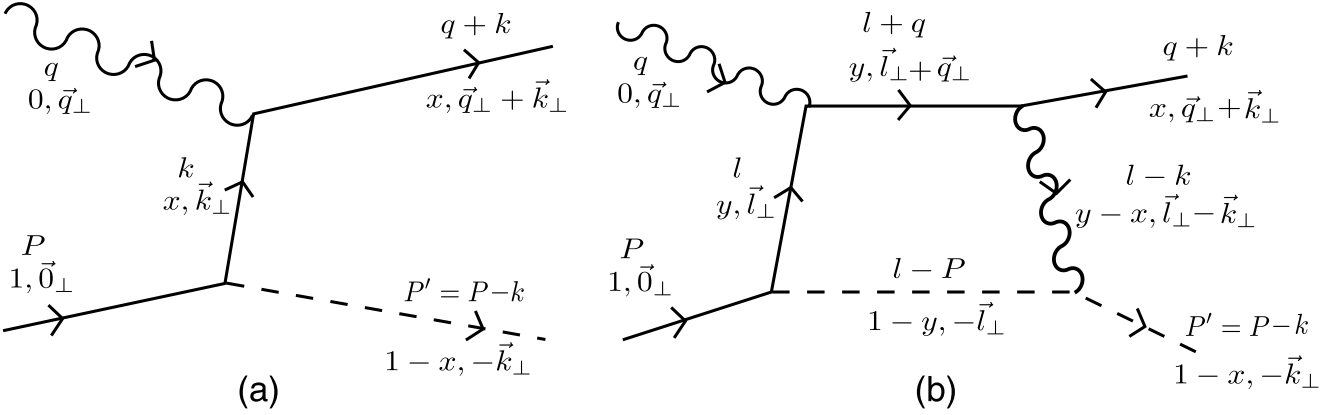}\\
\caption{\it (a) Tree level diagram and (b) diagram with final-state interaction.}
\label{fig:feynmandiagrams}
\end{center}
\end{figure}

In Ref. \cite{BHS} it was found that the
contributing amplitudes for $\gamma^* p \to q (qq)_0$ are given by the
following formulas through one loop order
which is depicted in Fig. \ref{fig:feynmandiagrams}:
\begin{eqnarray}
{\cal A}(\Uparrow \to \uparrow)&=&\frac{(m+xM)}{x}\ C\
(h+i{e_1e_2\over 8\pi}g_1)
\label{s1}\\
{\cal A}(\Uparrow \to \downarrow)&=&\ -\ \frac{(+k^1+{i} k^2)}{x }\ \ C\
(h+i{e_1e_2\over 8\pi}g_2)
\label{s3}\\
{\cal A}(\Downarrow \to \uparrow)&=&\ -\ \frac{(-k^1+{i} k^2)}{x }\ \ C\
(h+i{e_1e_2\over 8\pi}g_2)
\label{s2}\\
{\cal A}(\Downarrow \to \downarrow)&=&\frac{(m+xM)}{x}\ C\
(h+i{e_1e_2\over 8\pi}g_1) \ , \label{s4}
\end{eqnarray}
where
\begin{eqnarray}
C&=&-\ g\ e_1\ P^+\ {\sqrt{x}}\ 2\ x\ (1-x)
\label{s5}\\
h&=& {1\over {\vec k}_{\perp}^2+x
(1-x)(-M^2+{m^2\over x} +{\lambda^2\over 1-x})}\ ,
\label{s6}
\end{eqnarray}
and
\begin{eqnarray}
g_1&=&\int_0^1d\alpha\ {-1\over \alpha (1-\alpha){\vec k}_{\perp}^2
+\alpha \lambda_g^2 +(1-\alpha)B} \ ,
\label{s7}\\
g_2&=&\int_0^1d\alpha\ {-\alpha\over \alpha (1-\alpha){\vec
k}_{\perp}^2 +\alpha \lambda_g^2 +(1-\alpha)B}
\ .
\label{s8}
\end{eqnarray}
In the above, $e_1$ and $e_2$ are the quark and diquark charge,
and $M$, $m$, $\lambda$ and $\lambda_g$ are the nucleon, quark, diquark
and gluon mass, respectively.
Ref. \cite{BHS} fixed ${e_1e_2\over 4\pi}=-C_F\alpha_S$, where $C_F={4\over 3}$
in order to relate the above results to QCD.
We take $\lambda_g=0$ at the end of the calculation.
We note that the results (\ref{s7}) and (\ref{s8}) are for the
semi-inclusive deep inelastic scattering, and the results for the
Drell-Yan process have opposite signs compared to (\ref{s7}) and (\ref{s8})
\cite{Collins02,BHS2}.

The final-state interactions in semi-inclusive deep inelastic scattering
are commonly treated as a part of the proton distribution function
\cite{Collins02,BBH}.
If we adopt the same treatment for the wavefunctions,
we can consider that the final-state interactions
for the scalar diquark model depicted
in Fig. \ref{fig:feynmandiagrams} induce
the spin-dependent complex phases to the wavefunctions \cite{LC2008dsh}:
\begin{equation}
\left
\{ \begin{array}{l}
\psi^{\uparrow}_{+\frac{1}{2}} (x,{\vec k}_{\perp})=\frac{(m+xM)}{x}\,
\Big( 1+ia_1\Big)\, \varphi \ ,\\
\psi^{\uparrow}_{-\frac{1}{2}} (x,{\vec k}_{\perp})=
-\frac{(+k^1+{i} k^2)}{x }\,
\Big( 1+ia_2\Big)\, \varphi \ ,
\end{array}
\right.
\label{sn2n}
\end{equation}
\begin{equation}
\left
\{ \begin{array}{l}
\psi^{\downarrow}_{+\frac{1}{2}} (x,{\vec k}_{\perp})=
-\frac{(-k^1+{i} k^2)}{x }\,
\Big( 1+ia_2\Big)\, \varphi \ ,\\
\psi^{\downarrow}_{-\frac{1}{2}} (x,{\vec k}_{\perp})=\frac{(m+xM)}{x}\,
\Big( 1+ia_1\Big)\, \varphi \ ,
\end{array}
\right.
\label{sn2an}
\end{equation}
where
$a_1$ and $a_2$ are given by
\begin{equation}
a_{1,2}={e_1e_2\over 8\pi}\ ({\vec k}_{\perp}^2+B)\ g_{1,2}
\label{alpha12}
\end{equation}
with $g_{1,2}$ given in (\ref{s7}) and (\ref{s8}).

Using the wavefunctions (\ref{sn2n}) and (\ref{sn2an})
in the formulas (\ref{p11sa}), (\ref{p11sb}) and (\ref{p11sbh1pan}),
we obtain
\begin{eqnarray}
f_1(x,{\vec k}_{\perp})&=&
{1\over 16 \pi^3}\ \Big[ (M+\frac{m}{x})^2+{{\vec k}_{\perp}^2\over x^2}\Big]
\varphi^2\ ,
\label{f1xk}\\
f_{1T}^{\perp}(x,{\vec k}_{\perp})&=&
{1\over 16 \pi^3}\ 2\ {M\over x}\ (M+\frac{m}{x})\  \varphi^2\
{e_1e_2\over 8\pi}\ ({\vec k}_{\perp}^2+B)\ 
{1\over {\vec k}_{\perp}^2}\
{\rm ln}{({\vec k}_{\perp}^2 + B)\over B}\ ,
\label{f1Tpxk}\\
h_{1}^{\perp}(x,{\vec k}_{\perp})&=&
{1\over 16 \pi^3}\ 2\ {M\over x}\ (M+\frac{m}{x})\  \varphi^2\
{e_1e_2\over 8\pi}\ ({\vec k}_{\perp}^2+B)\
{1\over {\vec k}_{\perp}^2}\
{\rm ln}{({\vec k}_{\perp}^2 + B)\over B}\ .
\label{h1pxk}
\end{eqnarray}
The results in (\ref{f1Tpxk}) and (\ref{h1pxk}) agree with
those in Refs. \cite{BHS,BBH}
with an additional overall minus sign which should be corrected \cite{BH04}.

\subsection{Axial-Vector Diquark Model}

\begin{figure}[h!]
\begin{center}
\vspace{0.5cm}
\includegraphics[width=0.75\columnwidth]{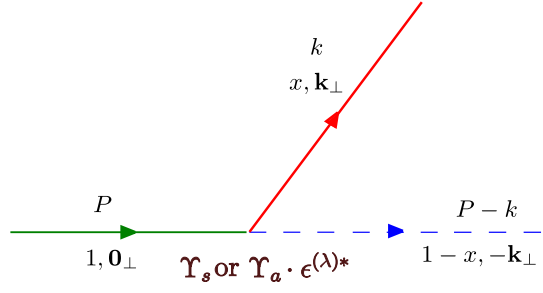}\\
\vspace{-1.0cm}
\caption{\it Diagram giving the light-cone wavefunctions of scalar and
axial-vector diquark models.}
\label{fig:waveftnsa}
\end{center}
\end{figure}

Jakob et al. \cite{JMR97} studied the scalar ($s$)
and axial-vector ($a$) diquark models
using the following nucleon-quark-diquark vertices:
\begin{equation}
\Upsilon_s=g_s(k^2)\ , \qquad
\Upsilon_a^\mu ={g_a(k^2)\over {\sqrt{3}}}\gamma^\mu\gamma_5 \ ,
\label{sa1aa}
\end{equation}
where $g_s(k^2)$ and $g_a(k^2)$ are form factors which we take as 1
in this paper for simplicity.
We can then obtain the light-cone wavefunctions of scalar and
axial-vector diquark models from Fig. \ref{fig:waveftnsa}.
In Ref. \cite{JMR97}, $\gamma_5(\gamma^\mu+{P^\mu\over M})$ appears at
the vertex of the axial-vector diquark model instead of just
$\gamma^\mu\gamma_5$ appearing in (\ref{sa1aa}).
However, (\ref{sa1aa}) is equivalent to the vertex of Ref. \cite{JMR97}
since ${P^\mu\over M}$ vanishes when the polarization sum
$\sum_{\lambda} {\epsilon}_{\mu}^{*(\lambda)} {\epsilon}_{\nu}^{(\lambda)}
= - g_{\mu\nu} + P_{\mu}P_{\nu}/M^2$, which Ref. \cite{JMR97} used,
is multiplied.

In order to obtain the light-cone wavefunctions of axial-vector diquark model,
we decompose the polarization sum
$\sum_{\lambda} {\epsilon}_{\mu}^{*(\lambda)} {\epsilon}_{\nu}^{(\lambda)}
= - g_{\mu\nu} + P_{\mu}P_{\nu}/M^2$ of Ref. \cite{JMR97}
to the following three polarization vectors ${\epsilon}^{{\mu}(\lambda)}$:
\begin{eqnarray}
{\epsilon}^{{\mu}(+1)}&=&
({\epsilon}^{0(+1)}, {\epsilon}^{1(+1)}, {\epsilon}^{2(+1)}, {\epsilon}^{3(+1)})
\ =\ {1 \over {\sqrt{2}}} (0,-1,-i,0),
\nonumber\\
{\epsilon}^{{\mu}(-1)}&=&{1 \over {\sqrt{2}}} (0,+1,-i,0),
\label{axipol}\\
{\epsilon}^{{\mu}(0)}&=&({P^3 \over M},0,0,{P^0 \over M}).
\nonumber
\end{eqnarray}
In order to calculate the effects of the final-state interactions,
we use for the gauge-field coupling to the axial-vector diquark
in Fig. \ref{fig:feynmandiagrams} the simple form
$i e_2\, g^{\alpha\beta}\, ((P-l)+(P-k))^\mu$, which is equivalent,
for each polarization state, to the gauge-field coupling to a scalar
diquark \cite{EHK08}.
We motivate this simple coupling by assuming that the QCD coupling to the diquark is
independent of the spin state of the diquark.

The two-particle Fock state for proton with $J^z = + {1\over 2}$ (positive helicity)
has six possible spin combinations for the quark and axial-vector diquark:
\begin{eqnarray}
\lefteqn{
\left|\Psi^{\uparrow}_{\rm two \ particle}(P^+, \vec P_\perp = \vec
0_\perp)\right> =
\int\frac{{\mathrm d} x \, {\mathrm d}^2
           {\vec k}_{\perp} }{\sqrt{x(1-x)}\, 16 \pi^3}
}
\label{vsn1}\\
&&
\left[ \ \ \,
\psi^{\uparrow}_{+\frac{1}{2}\, +1}(x,{\vec k}_{\perp})\,
\left| +\frac{1}{2}\, +1\, ;\,\, xP^+\, ,\,\, {\vec k}_{\perp}\right>
+\psi^{\uparrow}_{+\frac{1}{2}\, -1}(x,{\vec k}_{\perp})\,
\left| -\frac{1}{2}\, +1\, ;\,\, xP^+\, ,\,\, {\vec k}_{\perp}\right>
\right.
\nonumber\\
&&\left. {}
+\psi^{\uparrow}_{-\frac{1}{2}\, +1} (x,{\vec k}_{\perp})\,
\left| +\frac{1}{2}\, \ \ 0\, ;\,\, xP^+\, ,\,\, {\vec k}_{\perp}\right>
+\psi^{\uparrow}_{-\frac{1}{2}\, -1} (x,{\vec k}_{\perp})\,
\left| -\frac{1}{2}\, \ \ 0\, ;\,\, xP^+\, ,\,\, {\vec k}_{\perp}\right>
\right.
\nonumber\\
&&\left. {}
+\psi^{\uparrow}_{-\frac{1}{2}\, +1} (x,{\vec k}_{\perp})\,
\left| +\frac{1}{2}\, -1\, ;\,\, xP^+\, ,\,\, {\vec k}_{\perp}\right>
+\psi^{\uparrow}_{-\frac{1}{2}\, -1} (x,{\vec k}_{\perp})\,
\left| -\frac{1}{2}\, -1\, ;\,\, xP^+\, ,\,\, {\vec k}_{\perp}\right>\
\right] \ ,
\nonumber
\end{eqnarray}
where the two-particle states $|s_{\rm f}^z, s_{\rm b}^z; \ x, {\vec
k}_{\perp} \rangle$ are normalized as in (\ref{normalize}). $s_{\rm
f}^z$ and $s_{\rm b}^z$ denote the $z$-component of the spins of the
constituent fermion and boson, respectively, and the variables $x$ and
${\vec k}_{\perp}$ refer to the momentum of the fermion. The
wavefunctions are given by \cite{EHK08}
\begin{equation}
\left
\{ \begin{array}{l}
\psi^{\uparrow}_{+\frac{1}{2}\, +1} (x,{\vec k}_{\perp})=-{\sqrt{{2\over 3}}}
\ \frac{(-k^1+{i} k^2)}{x}\,
\Big( 1+ia_2\Big)\, \varphi \ ,\\
\psi^{\uparrow}_{-\frac{1}{2}\, +1} (x,{\vec k}_{\perp})=+{\sqrt{{2\over 3}}}
\ {(m+xM)\over x}\,
\Big( 1+ia_1\Big)\, \varphi \ ,\\
\psi^{\uparrow}_{+\frac{1}{2}\, \ 0} (x,{\vec k}_{\perp})=-{\sqrt{{1\over 3}}}
\ {(m+xM)\over x}\,
\Big( 1+ia_1\Big)\, \varphi \ ,\\
\psi^{\uparrow}_{-\frac{1}{2}\, \ 0} (x,{\vec k}_{\perp})=+{\sqrt{{1\over 3}}}
\ \frac{(+k^1+{i} k^2)}{x}\,
\Big( 1+ia_2\Big)\, \varphi \ ,\\
\psi^{\uparrow}_{+\frac{1}{2}\, -1} (x,{\vec k}_{\perp})=0\ ,\\
\psi^{\uparrow}_{-\frac{1}{2}\, -1} (x,{\vec k}_{\perp})=0\ ,
\end{array}
\right.
\label{vsn2}
\end{equation}
where the scalar part of the wavefunction
$\varphi (x,{\vec k}_{\perp})$ is given
in (\ref{wfdenom}).

Similarly, the wavefunctions for proton with $J^z = - {1\over 2}$ (negative helicity)
are given by \cite{EHK08}
\begin{equation}
\left
\{ \begin{array}{l}
\psi^{\downarrow}_{+\frac{1}{2}\, +1} (x,{\vec k}_{\perp})=0\ ,\\
\psi^{\downarrow}_{-\frac{1}{2}\, +1} (x,{\vec k}_{\perp})=0\ ,\\
\psi^{\downarrow}_{+\frac{1}{2}\, \ 0} (x,{\vec k}_{\perp})=-{\sqrt{{1\over 3}}}
\ \frac{(-k^1+{i} k^2)}{x}\,
\Big( 1+ia_2\Big)\, \varphi \ ,\\
\psi^{\downarrow}_{-\frac{1}{2}\, \ 0} (x,{\vec k}_{\perp})=+{\sqrt{{1\over 3}}}
\ {(m+xM)\over x}\,
\Big( 1+ia_1\Big)\, \varphi \ ,\\
\psi^{\downarrow}_{+\frac{1}{2}\, -1} (x,{\vec k}_{\perp})=-{\sqrt{{2\over 3}}}
\ {(m+xM)\over x}\,
\Big( 1+ia_1\Big)\, \varphi \ ,\\
\psi^{\downarrow}_{-\frac{1}{2}\, -1} (x,{\vec k}_{\perp})=+{\sqrt{{2\over 3}}}
\ \frac{(+k^1+{i} k^2)}{x}\,
\Big( 1+ia_2\Big)\, \varphi \ .
\end{array}
\right.
\label{vsn2a}
\end{equation}
In Eqs. (\ref{vsn2}) and (\ref{vsn2a}) $a_1$ and $a_2$ are given by (\ref{alpha12}),
and the coefficients of $\varphi$ are the matrix elements of
$\frac{\overline{u}(k^+,k^-,{\vec k}_{\perp})}{{\sqrt{k^+}}}
\gamma \cdot \epsilon^{*}
\frac{u (P^+,P^-,{\vec P}_{\perp})}{{\sqrt{P^+}}}$
which are
the numerators of the wavefunctions corresponding to
each constituent spin $s^z$ configuration.

Using the wavefunctions given in (\ref{vsn2}) and (\ref{vsn2a})
in the formulas (\ref{p11sa}) and (\ref{p11sb}),
we obtain
\begin{equation}
f_1(x,{\vec k}_{\perp})=
{1\over 16 \pi^3}\ \Big[ (M+\frac{m}{x})^2+{{\vec k}_{\perp}^2\over x^2}\Big]
\varphi^2\ ,
\label{f1xkavdq}
\end{equation}
\begin{equation}
f_{1T}^{\perp}(x,{\vec k}_{\perp})=-{1\over 3}
{1\over 16 \pi^3}\ 2\ {M\over x}\ (M+\frac{m}{x})\  \varphi^2\
{e_1e_2\over 8\pi}\ ({\vec k}_{\perp}^2+B)\
{1\over {\vec k}_{\perp}^2}\
{\rm ln}{({\vec k}_{\perp}^2 + B)\over B}\ .
\label{f1tpxkav}
\end{equation}
In the same way,
by using the wavefunctions given in (\ref{vsn2}) and (\ref{vsn2a})
in the formula (\ref{p11sbh1pan}),
we obtain the Boer-Mulders distribution function as
\begin{equation}
h_{1}^{\perp}(x,{\vec k}_{\perp})=
{1\over 16 \pi^3}\ 2\ {M\over x}\ (M+\frac{m}{x})\  \varphi^2\
{e_1e_2\over 8\pi}\ ({\vec k}_{\perp}^2+B)\
{1\over {\vec k}_{\perp}^2}\
{\rm ln}{({\vec k}_{\perp}^2 + B)\over B}\ .
\label{h1pxkav}
\end{equation}

The Sivers distribution function $f_{1T}^\perp$
is given by the overlaps of the proton wavefunctions
of positive and negative helicities \cite{BHS,BBH}.
As we can see in (\ref{vsn2}) and (\ref{vsn2a}), only the wavefunctions with
$s_{\rm b}^z = 0$ contribute to the overlaps for $f_{1T}^\perp$.
We find that the wavefunctions with $s_{\rm b}^z = 0$ in (\ref{vsn2}) and (\ref{vsn2a})
have the exactly same structures as the wavefunctions for the quark and scalar diquark system
given in (\ref{sn2n}) and (\ref{sn2an}), except for the overall constant factor
$-{1\over {\sqrt{3}}}$ for the positive helicity wavefunctions and
$+{1\over {\sqrt{3}}}$ for the negative helicity ones.
These constant factors can be understood by the Clebsch-Gordan coefficients for
the combination of spin ${1 \over 2}$ and spin $1$ states \cite{EHK08}.
Therefore, the overlaps of positive and negative helicities for the quark and axial-vector
diquark system have an overall factor of $-{1\over 3}$ compared to the
overlaps of (\ref{sn2n}) and (\ref{sn2an}) for the quark and scalar diquark system.
This is the reason why, in the diquark models with which we work in this paper, we have the same
Sivers distribution functions for the cases of the scalar diquark and axial-vector diquark,
except for the difference of the additional overall constant factor
$-{1\over 3}$ of the axial-vector diquark model \cite{EHK08},
as we can see in (\ref{f1Tpxk}) and (\ref{f1tpxkav}).

On the other hand, the Boer-Mulders distribution function $h_{1}^{\perp}$
is given by the overlaps of the wavefunctions of opposite quark spin states within
a given proton helicity state.
Concerning the two quark spin states for a fixed $s_{\rm b}^z$ value
within a given proton helicity state,
the wavefunctions in (\ref{vsn2}) and (\ref{vsn2a})
have the same structures as the wavefunctions of the scalar diquark model
given in (\ref{sn2n}) and (\ref{sn2an}).
This structure is given by the relation between the light-cone and Bjorken-Drell spinors
\cite{EHK08}.
Therefore, the Boer-Mulders distribution of the axial-vector diquark model is the same
as that of the scalar diquark model as we can see in (\ref{h1pxk}) and (\ref{h1pxkav}).
We summarize the results in Table 1.
We note that Ref. \cite{burkardt08} studied the Boer-Mulders distribution
functions $h_1^\perp$ for a variety of phenomenological models and found that
the signs of $h_1^\perp$ are all negative for the models which they studied.

\begingroup
\begin{table}[t]
\vspace*{0.6cm}
\label{tab:eigenvalue}
\begin{center}
\begin{tabular}{|c|c|c|}
\hline\hline
 &\ \ \ \ \ Sivers Function\ \ \ \ \ &Boer-Mulders Function\\
\hline\hline
\ \ Scalar Diqurrk Model \ \ & \ \ \ \ \ -1 \ \ \ \ \ &
\ \ \ \ \ -1 \ \ \ \ \  \\
\hline
\ \ Axial-Vector Diquark Model \ \ & \ \ \ \ \ $+{1\over 3}$ \ \ \ \ \ &
\ \ \ \ \ -1 \ \ \ \ \  \\
\hline\hline
\end{tabular}
\end{center}
\vspace*{-0.5cm}
\caption{Relative signs and magnitudes of Sivers and Boer-Mulders distribution
functions in scalar and axial-vector diquark models.}
\end{table}
\endgroup

\section{Conclusion}

In this paper we find the light-cone wavefunction representations of
the Sivers and Boer-Mulders distribution functions.
A necessary condition for the existence of these representations is that
the light-cone wavefunctions have complex phases.
We induce the complex phases by incorporating the final-state interactions
into the light-cone wavefunctions in the scalar and axial-vector diquark models,
and then we calculate explicitly the Sivers and Boer-Mulders distribution
functions from the light-cone wavefunction representations.
The results are the same as those obtained
from the direct calculation of the hadronic
tensor without employing the concept of the light-cone wavefunction,
since the essential interpretation of the final-state interaction is
identical in both calculations.
However, the analysis in this paper by using the light-cone wavefunction
representations is useful for understanding the natures of the Sivers and Boer-Mulders
distribution functions in a systematic way.
In the light-cone wavefunction representations
the Sivers distribution function is given by the overlap of the wavefunctions of the same
quark spin states of the opposite nucleon spin states,
whereas the Boer-Mulders distribution function
is given by the overlap of the wavefunctions of the opposite quark spin states
within a given nucleon spin state.
From these properties of the light-cone wavefunction representations,
we can understand why the Sivers distribution function has the opposite signs
with the factor 3 difference in magnitude for the scalar and axial-vector diquark models,
whereas the Boer-Mulders distribution function has the same sign
and magnitude for these diquark models.

\section*{Acknowledgments}
This work was supported in part by the International Cooperation
Program of the KICOS (Korea Foundation for International Cooperation
of Science \& Technology),
and by the Korea Research Foundation Grant funded by the Korean
Government
(KRF-2008-313-C00166).


\begin{thebibliography}{99}

\bibitem{BHS} S.J. Brodsky, D.S. Hwang, and I. Schmidt,
Phys. Lett. B {\bf 530}, 99 (2002).

\bibitem{Sivers}
  D.W.~Sivers,
  Phys.\ Rev.\  D {\bf 41}, 83 (1990);
  Phys.\ Rev.\  D {\bf 43}, 261 (1991).

\bibitem{Collins02} J.C. Collins, Phys. Lett. B {\bf 536}, 43 (2002).

\bibitem{JY} X. Ji and F. Yuan, Phys. Lett. B {\bf 543}, 66 (2002).

\bibitem{BJY}
A. Belitsky, X. Ji, and F. Yuan, Nucl. Phys. B {\bf 656}, 165 (2003).

\bibitem{BBH} D. Boer, S.J. Brodsky, and D.S. Hwang,
Phys. Rev. D {\bf 67}, 054003 (2003).

\bibitem{BMP} D. Boer, P.J. Mulders, and F. Pijlman,
Nucl. Phys. B {\bf 667}, 201 (2003).

\bibitem{BM98} D. Boer and P.J. Mulders,
Phys. Rev. D {\bf 57}, 5780 (1998).

\bibitem{BL} G.P. Lepage and S.J. Brodsky,
Phys. Rev. D {\bf 22}, 2157 (1980).

\bibitem{BPP} S.J. Brodsky, H.C. Pauli, and S.S. Pinsky,
Phys. Rep. {\bf 301}, 299 (1998).

\bibitem{BD80} S.J. Brodsky and S.D. Drell,
Phys. Rev. D {\bf 22}, 2236 (1980).

\bibitem{BH99} S.J. Brodsky and D.S. Hwang,
Nucl. Phys. B {\bf 543}, 239 (1999).

\bibitem{BHMS} S.J. Brodsky, D.S. Hwang, B.-Q. Ma, and I. Schmidt,
Nucl. Phys. B {\bf 593}, 311 (2001).

\bibitem{DFJK} M. Diehl, T. Feldmann, R. Jakob, and P. Kroll,
Nucl. Phys. B {\bf 596}, 33 (2001), Erratum-ibid. B {\bf 605}, 647 (2001).

\bibitem{BDH} S.J. Brodsky, M. Diehl, and D.S. Hwang,
Nucl. Phys. B {\bf 596}, 99 (2001).

\bibitem{BGH06} S.J. Brodsky, S. Gardner, and D.S. Hwang,
Phys. Rev. D {\bf 73}, 036007 (2006).

\bibitem{BHS2} S.J. Brodsky, D.S. Hwang, and I. Schmidt,
Nucl. Phys. B {\bf 642}, 344 (2002).

\bibitem{LC2008dsh} D.S. Hwang, PoS (LC2008) 033;
{\it Proceedings of LIGHT CONE 2008 Relativistic Nuclear and Particle Physics}
(July 7-11, 2008, Mulhouse, France).

\bibitem{BH04} M. Burkardt and D.S. Hwang, Phys. Rev. D {\bf 69}, 074032 (2004).

\bibitem{JMR97} R. Jakob, P.J. Mulders, and J. Rodrigues,
Nucl. Phys. A {\bf 626}, 937 (1997).

\bibitem{EHK08} J. Ellis, D.S. Hwang, and A. Kotzinian,
Phys. Rev. D {\bf 80}, 074033 (2009).

\bibitem{burkardt08} M. Burkardt and B. Hannafious,
Phys. Lett. B {\bf 658}, 130 (2008).

%

\end{thebibliography}
\end{document}